\documentstyle[prl,twocolumn,aps]{revtex}
\begin{document}
\draft
\title{One-Electron Spectral Functions of \\
the Attractive Hubbard Model for
Intermediate Coupling}
\author{Maxim Yu. Kagan$^{(a)}$, Raymond Fr\'{e}sard$^{(b)}$, 
Massimiliano Capezzali$^{(b)}$ and Hans Beck$^{(b)}$}
\address{$^{(a)}$  P.L. Kapitza Institute for Physical Problems, Kosygin 
Str. 2,
117334 Moscow, Russian Federation\\
$^{(b)}$ Institut de Physique, Universit\'{e} de Neuch\^{a}tel, Rue A.L.
Breguet 1,
2000 Neuch\^{a}tel, Switzerland}
%
\twocolumn[
\maketitle
\widetext

\vspace*{-1.0truecm}

\begin{abstract}
\begin{center}
\parbox{14cm}{We calculate the one-electron spectral function of the 
attractive 
(negative-$U$)
Hubbard model. We work in the intermediate coupling and low density regime and
obtain the self-energy in an approximate analytical form. The excitation 
spectrum is found to 
consist of three branches. The results are obtained in a framework, based
on the self-consistent T-matrix approximation, which is compatible with the
Mermin-Wagner theorem.}
\end{center}
\end{abstract}
\pacs{
\hspace{1.9cm}
}]
\narrowtext
\section*{(a) Introduction}
The Hubbard model involving electrons on a lattice, subject to an attractive interaction when
they are on the same site, is one of the simplest models for describing superconductivity.
Despite of its simplicity, it has turned out to be very challenging for the theoreticians
to give a simple description of its properties which is valid in the various regimes 
of coupling strength.
In the weak coupling regime, the link with BCS theory of superconductivity has 
been done by Nozi\`{e}res and Schmitt-Rink \cite{nozi1}. At sufficiently low $T$, an
instability of the Fermi sea towards superconductivity occurs. In three dimensions,
the transition is essentially mean-field in character. In the opposite strong coupling
limit ($|U|\rightarrow\infty$), the electrons form bound pairs which are immobile since they
can only move via virtual ionization with an infinite energy barrier. However, for large
but finite $U$, those bound pairs essentially behave like heavy hard core bosons (with
an effective mass $m^{*}\sim m{U\over t}$) which are undergoing Bose-Einstein 
condensation at sufficiently low $T$. On the lattice, $T_{C}$ vanishes in the
limit $|U|\rightarrow\infty$, while in the continuum limit, it remains finite 
\cite{Zwerger}. This difference is due to the absence of a pair hopping term when working
on the lattice.\\
In the intermediate coupling regime, the physics will be dominated by the interplay between the
quasi-particles and the bound pairs, which may lead to non-trivial behavior.\\
Even though it is still lacking a microscopical derivation, the model is 
interesting
in its own right, since it allows for studying various routes leading to 
superconductivity.
Since the interaction is local, it will be $s$-wave superconductivity, but 
the generalization
to non-local interaction can be considered \cite{meintrup}. In the weak 
coupling regime,
perturbation theory is expected to work, and this has been worked out by 
a series 
of authors \cite{galitski}, some of them focusing on 2-d systems
\cite{randeria}.  Of special interest is the low-density regime where 
chances of obtaining 
meaningful results are better, since the ratio of the scattering length 
to the average inter-particle 
distance can be used as a small parameter. Unfortunately those calculations are
quickly becoming very involved since the simplest conserving approximation is
the self-consistent T-matrix approach 
\cite{fresard,rodriguez,haussmann}. Alternatively, 
Variational-Monte-Carlo (VMC) 
calculations, based on the Gutzwiller wave-function \cite{dent} and 
Quantum-Monte-Carlo (QMC) simulations have been performed \cite{singer}. 
These methods
are providing results which then generate a need for a qualitative
analytical understanding. To that aim simpler calculations based on 
Hubbard-Stratonovitch
decoupling of the interaction \cite{morten}, slave-boson mean-field 
calculations (see for instance Ref. \cite{bulka} and
references therein), or on the moment calculation of
the electronic spectral function have been performed \cite{micnas}. 
Unfortunately the latter does not account for the damping of the 
quasi-particles.\\
The aim of this paper is to treat analytically the intermediate coupling 
regime, which is the
most delicate. This allows us to give
an analytical account of the results obtained with QMC simulations. We first
review the self-consistent T-matrix approximation. 
As pointed out by several authors \cite{schafroth}, the corresponding 
numerical calculations
typically yield a superconducting instability at a finite $T$, even in 
two dimensions. This
contradicts the Mermin-Wagner theorem. We then propose an alternative 
scheme which
complies with this theorem. 
We then proceed to the calculation of the electronic
structure.
\section*{(b) Theoretical framework}
We study the Hubbard model on the square lattice :
\begin{equation}\label{ham}
H=\sum_{i,j}^{}{\sum_{\left\langle \sigma \right\rangle}^{}{t_{ij}
c^{\dagger}_{i,\sigma}c_{j,\sigma}}}+U\sum_{i}{}{n_{i,\uparrow}n_{i,\downarrow}}.
\end{equation}
We consider an attractive interaction ($U<0$) in the
intermediate coupling regime ($|U|{\ \lower-1.2pt\vbox{\hbox{\rlap{$<$}\lower5pt\vbox{\hbox{$\sim$}}}}\ }W$), $W$ being the
band width. In 2 dimensions, any attractive potential has a bound state. In the
case $|U|=W$, the binding energy $E_{b}$ has been found to be $E_{b}\approx
0.2W$ \cite{rodriguez}, namely $E_{b}\ll W$. We are thus in a situation where bound pairs
exist and have a strong influence on the physics via the splitting of the non-interacting
band into 2 sub-bands. In this regime, the pairs are extended. They become purely local
only in the $|U|=\infty$ limit since for any finite $U$ they can move via virtual
ionization \cite{nozi1}. We also note that the BCS theory successfully 
describes
the weak coupling regime. However there does not exist any analytical theory
in the intermediate coupling regime, and most results are obtained out of 
numerical
simulations \cite{randeria,singer}. In the low-density regime, the 
self-consistent
T-matrix approximation is expected to be exact and has been solved by a variety
of authors \cite{fresard,rodriguez,haussmann}. Unfortunately, numerical 
difficulties
prevented those authors from obtaining results for arbitrary $U$.  We also
note that the numerical solutions may lead to unphysical results such
as a finite critical temperature for Bose condensation of
the pairs in two dimensions, which is contradicting the Mermin-Wagner theorem.
We believe (see below) that this is due to the
use of an inappropriate expression for the particle density. That however 
does not discredit
the scheme, and we are basing our approach on it. It amounts to solving
\begin{equation}\label{tfrai}
T(\vec{q},i\nu_{n})={-U\over 1+
U\chi(\vec{q},i\nu_{n})} \;\;\;\;\;\;\;\;\;\;\;\;\;\;\;\;\;\;\;\;\;\;\;\;\;\;\;\;\;\;\;\;\;\;\;\,\;
\end{equation}
\begin{equation}\label{chi}
\chi(\vec{q},i\nu_{n})=\beta^{-1}\sum_{\vec{p},i\omega_{n}}^{}{
G(\vec{p},i\omega_{n})G(\vec{q}-\vec{p},i\nu_{n}-i\omega_{n})} \;
\end{equation}
\begin{equation}\label{s1tot}
\Sigma(\vec{q},i\omega_{n})=-\beta^{-1}\sum_{\vec{p},i\nu_{n}}^{}{
T(\vec{p},i\nu_{n})G(\vec{p}-\vec{q},i\nu_{n}-i\omega_{n})}
\end{equation}
\begin{equation}\label{g1tot}
G(\vec{q},i\omega_{n})={1\over i\omega_{n}-t_{\vec{q}}+\mu-
\Sigma(\vec{q},i\omega_{n})}\, .\;\;\;\;\;\;\;\;\;\;\;\;\;\;\;\;\;\;\,\;
\end{equation}
Here, $\omega_{n}$ are Fermionic, and $\nu_{n}$ Bosonic Matsubara frequencies.
This set of equations is valid {\it above} $T_{C}$, as no anomalous Green's function
enters.  Otherwise one can resort to the scheme obtained by Pedersen {\it et al.},
by functional derivative techniques \cite{morten}. This approximation is conserving
and it diagrammatically corresponds to summing up the dressed particle-particle 
ladder which includes the leading order in an expansion
in $k_{F}a$ \cite{galitski,haussmann}. 
Another important quantity is the two-particle Green's function which is defined by :
\begin{eqnarray}\label{g2b}
G^{(2)}(\vec{q},i\nu_{n})=\int_{0}^{\beta}{e^{i\nu_{n}\tau}
\left\langle T_{\tau}\left[ Q(\vec{q},\tau)Q^{\dagger}(-\vec{q},0) \right] \right\rangle d\tau},
\end{eqnarray}
where $T_{\tau}$ is the usual time-ordering operator and the operator
\begin{eqnarray}\label{creaop}
Q^{+}(\vec{q})={1\over N}\sum_{\vec{k}}^{}{c_{-\vec{k},\uparrow}^{\dagger}
c_{\vec{k}-\vec{q},\downarrow}^{\dagger}}
\end{eqnarray}
creates a pair having (center-of-mass) wave vector $\vec{q}$. 
$G^{(2)}(\vec{q},i\nu_{n})$ is related to the T-matrix by :
\begin{eqnarray}\label{g2}
G^{(2)}(\vec{q},i\nu_{n})={U+T(\vec{q},i\nu_{n})\over U^{2}}.
\end{eqnarray}
We calculate $G^{(2)}(\vec{q},i\nu_{n})$ by inserting the free-electron Green's
function into expression (\ref{chi}) for $\chi(\vec{q},i\nu_{n})$. For simplicity,
we approximate the density of states (DOS), $\rho(\epsilon)$, of the tight-binding 
band resulting from the Hamiltonian (\ref{ham}) by
the square DOS (i.e. $\rho(\epsilon)={1\over W}$ for
$|\epsilon|\leq{W\over 2}$ and $\rho(\epsilon)=0$ otherwise).\\ 
For small momenta, $G^{(2)}(\vec{q},i\nu_{n})$ is given by :
\begin{eqnarray}\label{gexp}
G^{(2)}(\vec{q},i\nu_{n})={1\over 2W\left(1-{q^{2}\over 16} \right)}\times
\;\;\;\;\;\;\;\;\;\;\;\;\;\;\;\;\;\;\;\;\;\; \nonumber \\
{\ln{\left( {i\nu_{n}+\mu_{B}-|E_{b}|-2W+{q^{2}t\over 2}\over
i\nu_{n}+\mu_{B}-|E_{b}|-{q^{2}t\over 2}} \right)}\ln{\Phi}
\over
\ln{\Phi}-\left(1+{q^{2}\over 16} \right)
\ln{\left( {i\nu_{n}+\mu_{B}-|E_{b}|-2W+{q^{2}t\over 2}\over
i\nu_{n}+\mu_{B}-|E_{b}|-{q^{2}t\over 2}}\right)}},
\end{eqnarray}
where $\mu_{B}=2\mu+W+|E_{b}|$, $\Phi={2W+|E_{b}|\over |E_{b}|}$ and $|E_{b}|$
is the binding energy of a pair. 
The binding energy is obtained
as a solution of  :
\begin{equation}\label{bind}
-{1\over U}=\chi(\vec{q}=\vec{0}, \omega=E_{b})|_{\mu=-{W\over 2}},
\end{equation}
which yields 
\begin{equation}\label{bind2}
|E_{b}|=2W\left( {1\over e^{{-2W\over U}}-1} \right).
\end{equation}
The form (\ref{gexp}) has the correct behavior for $\nu_{n}$ going to infinity in the
low density regime, i.e.
$G^{(2)}(\vec{q},i\nu_{n})\rightarrow{1\over i\nu_{n}}$. \\
The spectrum of 
$G^{(2)}(\vec{q},i\nu_{n})$ presents two 
features : (i) a sharp quasi-particle peak, which can be found by expanding (\ref{gexp}) 
with respect to $i\nu_{n}+\mu_{B}-{q^{2}t\over 2}$; (ii) a continuous spectrum which extends
over energies above the one of the quasi-particle. Correspondingly, the lowest order form of the
T-matrix, valid for small wave vector and frequency is given by :
\begin{equation}\label{t0}
T_{0}(\vec{q},i\nu_{n})={-|E_{b}|^{2}\Phi\over i\nu_{n}-{\vec{q}\,^{2}\over 
4m_{0}^{*}}+\mu_{B}}.
\end{equation}
The mass renormalization factor of a pair is given by :
\begin{eqnarray}\label{renorm}
Z\equiv{m\over m_{0}^{*}}={W+|E_{b}|\over W}-{|E_{b}|^{2}\over 2W^{2}}\Phi\ln{\Phi}.
\end{eqnarray}
In the intermediate coupling
regime, the mass is only weakly renormalized while in the strong coupling regime, there
is a strong renormalization of the order ${W\over|U|}$.
Due to the relationship between the two-particle T-matrix and the two-particle
Green's function, the quantity $\mu_{B}$ that we defined above does represent the chemical 
potential of a pair, which has bosonic character. \\
For $\vec{q}$ 's close to the nesting vector $\vec{Q}=(\pi,\pi)$, we obtain:
\begin{equation}\label{teta}
T_{0}(\vec{q},i\nu_{n})={-U^{2}\over i\nu_{n}+{(\vec{q}-\vec{Q})^{2}\over 4m^{*}}+2\mu+|U|}.
\end{equation}
In the vicinity of the zone corner, the renormalization of the pair-mass is different from
the one close to the zone center. Even in the intermediate coupling regime, it is
strongly renormalized to be ${m^{*}\over m}\approx{|U|\over t}$. 
At $\vec{q}=\vec{Q}$, the form (9) of the T-matrix is actually exact, related to the fact that the
creation operator
\begin{equation}
\eta^{\dagger}=\sum_{\vec{p}}^{}{c_{\vec{p}+\vec{Q}, \uparrow}^{\dagger}
c_{-\vec{p}, \downarrow}^{\dagger}}
\end{equation}
of an "$\eta$-pair" with center of mass momentum $\vec{Q}$, satisfies the simple commutation
relation \cite{yang,nowak,demler} :
\begin{equation}
\left[ H,\eta^{\dagger} \right]=(U-2\mu)\eta^{\dagger}.
\end{equation}
Using the above found expressions (\ref{t0}) and (\ref{teta}), 
we can calculate the self-energy.
To lowest order, we insert the free-electron Green's function in Eqn. (\ref{s1tot}).
The first contribution to the self-energy arises from the poles of the T-matrix.
Due to the statistical factors we obtain (to that order of approximation) that the contribution of
the $\eta$-resonance is exponentially small, as well as those following from the poles
of the Green's function. After performing analytical continuation, we are left with :
\begin{equation}
\Sigma_{1}(\vec{k},\omega)={U^{2}n_{d}\over\omega+t_{\vec{k}}
-\mu+\mu_{B}+i0^{+}}.
\end{equation}
The quantity $n_{d}$ will be defined below, in Eqns. (\ref{doubleocc}) and
(\ref{doubleocc2}). 
$\Sigma_{1}(\vec{k},\omega)$ yields then the Green's function as :
\begin{eqnarray}\label{g2p}
G(\vec{k},\omega)=\;\;\;\;\;\;\;\;\;\;\;\;\;\;\;\;\;\;\;\;\;\;\;\;\;\;\;\;\;\;\;\;
\;\;\;\;\;\;\;\;\;\;\;\;\;\;\;\;\;\;\;\;\;\;\;\;\;\;\;\;\;\;\;\;\nonumber \\
{1\over 2}\left(1+
{2\left(t_{\vec{k}}-\mu \right)-\mu_{B}\over
x_{\vec{k}}}\right)
{1\over\omega+{1\over 2}\mu_{B}-{1\over 2}x_{\vec{k}}+i0^{+}} \;\,\nonumber \\
+{1\over 2}\left(1-{2\left(t_{\vec{k}}-\mu \right)+\mu_{B}\over
x_{\vec{k}}}\right)
{1\over\omega+{1\over 2}\mu_{B}+{1\over 2}x_{\vec{k}}+i0^{+}}, 
\end{eqnarray}
where $x_{\vec{k}}=\sqrt{\left( 2\left( t_{\vec{k}}-\mu \right)+\mu_{B} \right)^2
+4U^{2}n_{d}}$. We  
immediately note the two limiting behaviors, with respect to momentum $\vec{k}$ :
\begin{equation} 
x_{\vec{k}}\approx \Delta
+2t\gamma k^{2}, 
\end{equation}
with $\gamma={|E_{b}|\over \Delta}$ and 
\begin{eqnarray}
\Delta=\sqrt{|E_{b}|^2+4U^{2}n_{d}}
\end{eqnarray}
for small momenta; respectively, 
\begin{equation} 
x_{\vec{k}}\approx 2\left( t_{\vec{k}}-\mu \right)+\mu_{B}+
{2U^{2}n_{d}\over 2\left( t_{\vec{k}}-\mu \right)+\mu_{B}},
\end{equation}
for large momenta. \\
At this stage of the calculation, the Green's function has a two-pole structure. 
The lower excitation branch corresponds to quasi-bound
fermions (hereafter denoted as "bosonic" band), while the upper 
branch describes the unpaired fermions (fermionic band). 
At small
momenta, we obtain:
\begin{eqnarray}\label{g111}
G(\vec{k},\omega)={\Delta+|E_{b}|\over 2\Delta}
{1\over \omega+{1\over 2}\left( \mu_{B}-\Delta \right)-
\gamma tk^{2}+i0^{+}} \;\;\;\;\;\nonumber \\
+{\Delta-|E_{b}|\over 2\Delta}
{1\over \omega+{1\over 2}\left( \mu_{B}+\Delta \right)+\gamma tk^{2}+i0^{+}}\;\;\;\;\;\;
\end{eqnarray}
with the spectral weight mainly located in the unpaired fermion band 
(first contribution in Eqn. (\ref{g111})). At large momenta, 
the Green's function results into :
\begin{eqnarray}
G(\vec{k},\omega)={1-{2U^{2}n_{d}
\over \left(2\left(t_{\vec{k}}-\mu \right)+\mu_{B} \right)^{2}}
\over \omega-\left( t_{\vec{k}}-\mu \right)+i0^{+}} \;\;\;\;\;\;\;\nonumber \\
+{{2U^{2}n_{d} 
\over \left(2\left(t_{\vec{k}}-\mu \right)+\mu_{B} \right)^{2}}
\over \omega+t_{\vec{k}}-\mu+\mu_{B}+i0^{+}}, \;
\end{eqnarray}
where the weight of the paired fermion band is even smaller than for small momenta.
The form of the Green's function Eqn. (\ref{g2p}) 
differs from the one of Ref.\cite{vilk2} because
the chemical potential $\mu$ is located below the fermionic band in our problem.\\
We note that there are two equivalent expressions for the particle density 
operator $\hat{n}$:
\begin{equation}\label{equiv}
\hat{n}_{i}=\sum_{\sigma}^{}{\left(\hat{n}_{i,\sigma}(1-\hat{n}_{i,-\sigma})+
\hat{n}_{i,\sigma}\hat{n}_{i,-\sigma}\right)}
\end{equation}
On one hand, we can use the left hand side to express the particle density $n$ as
$n_{1}$, where the subscript 1 indicates that the density is calculated out
of the one-particle Green's function :
\begin{equation}\label{densi}
n_{1}=\beta^{-1}\sum_{i\omega_{n},\vec{k}}^{}{\sum_{\sigma}^{}{
G_{\sigma}(\vec{k},i\omega_{n})e^{i\omega_{n}0^{+}}}} \quad .
\end{equation}
Alternatively, we may use the r.h.s. of Eqn. (\ref{equiv}) 
by separating explicitly the contributions from
the 
unpaired fermions (first term) and the doubly occupied sites (second term). The total
density $n_{d}$ of the latter is given by
\begin{eqnarray}\label{doubleocc}
n_{d}={1\over \beta}\sum_{\vec{q}}^{}{\sum_{i\nu_{n}}^{}{
G^{(2)}(\vec{q},i\nu_{n})e^{i\nu_{n}0^{+}}
}} \;\;\;\;\;\;\;\;\;\;\;\;\;\;\;\;\;\;\;\;\;\;\;\;\nonumber \\
=\sum_{\vec{q}}^{}{\int_{-\infty}^{\infty}{{d\omega\over 2\pi}Im\left\{ 
G^{(2)}(\vec{q},\omega+i0^{+}) \right\}N_{B}(\omega)
}},
\end{eqnarray}
$N_{B}(\omega)$ being the Bose-Einstein distribution function. Owing to the latter, at
low temperatures, only the low-energy part of the two-particle spectrum, 
$Im\left\{ 
G^{(2)}(\vec{q},\omega+i0^{+}) \right\}$, will contribute to $n_{d}$. According to
the discussion following Eqn. (\ref{bind2}), this low-energy
part has two contributions : the sharp quasi-particle excitation, given by 
(\ref{t0}), which dominates for small momenta, and the low-frequency tail of the continuous 
spectrum of $G^{(2)}(\vec{q},i\nu_{n})$. Since it does not have a sharp structure, it would
only contribute a featureless background to the one-electron spectral function;
thus, we have neglected it for the calculation of the self-energy 
$\Sigma_{1}(\vec{k},\omega)$. 
Neglecting the continuum also in calculating $n_{d}$, we find, according to Eqns. 
(\ref{g2}), (\ref{t0}) and (\ref{doubleocc}) :
\begin{eqnarray}\label{doubleocc2}
n_{d}=R\sum_{\vec{q}}^{}{N_{B}\left( {q^{2}\over 4m}Z-\mu_{B} \right)
}.
\end{eqnarray} 
Thus, in our approximation, the number of doubly occupied sites $n_{d}$ is given by
the "number of bosons", $n_{B}$
\begin{equation}\label{bosons}
n_{B}\equiv\sum_{\vec{q}}^{}{N_{B}\left( {q^{2}\over 4m}Z-\mu_{B} \right)
},
\end{equation}
weighted by a factor 
\begin{eqnarray}\label{residue}
R=\left( {|E_{b}|\over 2W} \right)^{2}\Phi\left( \ln{\Phi} \right)^{2},
\end{eqnarray}
which is the residue of the two-particle
Green's function at the bottom of the two-particle band.
These bosons, having energy $E_{B}={q^{2}\over 4m}Z$ and chemical potential
$\mu_{B}$, represent pairs of electrons being (virtually) bound by the on-site
attraction. These results correspond to the observation of other authors 
(see, for example Ref. \cite{haussmann}) that, for sufficiently strong attraction, the
two-particle Green's function, (respectively the T-matrix) can be interpreted as a 
"bosonic Green's function". However, there is a weight factor
between the two. For large $|U|$, this weight factor $R$ goes to one : in the strong coupling
limit, all the double occupancy is due to 
coherently propagating quasi-bound pairs.
In the intermediate coupling regime, which we want to consider 
($|U| {\ \lower-1.2pt\vbox{\hbox{\rlap{$<$}\lower5pt\vbox{\hbox{$\sim$}}}}\ }W$), the
weight factor $R$ is roughly 0.5 : only about one half of the two-particle spectrum in
Eqn. (\ref{gexp}) is resulting from coherent excitations.
In the weak coupling regime,
$R$ vanishes, thus rendering our approximation invalid in this limit. \\
Since, in our approach, 
there is gap  in the one-electron spectrum,
separating the "bosonic" band from  the fermionic one,  
there
is no difficulty to obtain the number of unpaired electrons $n_{F}$ as :
\begin{equation}\label{densf}
n_{F}=\sum_{\vec{k},\sigma}^{}{\Xi_{\vec{k}}f_{F}
(\varepsilon_{\vec{k},\sigma})},
\end{equation}
where $f_{F}(\varepsilon_{\vec{k},\sigma})$ is the usual Fermi distribution function.
The dispersion $\varepsilon_{\vec{k},\sigma}$ of the unpaired fermions and 
the spectral weight $\Xi_{\vec{k}}$ entering Eqn. (\ref{densf}) are :
\begin{eqnarray} \label{gksi}
\varepsilon_{\vec{k},\sigma}={1\over 2}\left( x_{\vec{k}}-\mu_{B} \right), 
\;\;\;\;\;\;\;\;\;\;\;\;\;\;\;\;\;\;\;\; \nonumber \\
\Xi_{\vec{k}}={1\over 2}\left(1+{2\left( t_{\vec{k}}-\mu \right)-\mu_{B}\over x_{\vec{k}}} \right).
\end{eqnarray}
The total particle density 
results into :
\begin{equation}\label{densto}
n=n_{F}+2n_{d}.
\end{equation}
For non-interacting particles,
we may equally well use both r.h.s. and l.h.s of Eqn. (\ref{equiv}) to calculate the density,
since Wick's theorem applies.
However, the identity (\ref{equiv}) may be
violated in an approximate treatment like perturbation theory. This is the reason why
the self-consistent T-matrix calculation breaks down at low $T$ in two dimensions.
By calculating the density by means of Eqn. (\ref{densi}), there is nothing
preventing $T(\vec{q}=\vec{0},\omega=0)$ from diverging at finite $T$,
signaling a phase transition, while using Eqn. (\ref{densto}) 
would definitively keep $T(\vec{0},0)$
finite for any finite temperature. By making use of Eqn. (\ref{densto}), we 
make sure that
Bose condensation can only take place at $T=0$, in agreement with the 
Mermin-Wagner theorem. Indeed, 
according to the expressions (\ref{doubleocc2}) and (\ref{bosons}) for $n_{d}$ and
$n_{B}$, respectively, the bosonic chemical potential $\mu_{B}$, for a given
$n_{d}$ (or $n_{B}$) and for $d=2$, is different from zero at any finite
temperature, which inhibits Bose condensation, except for $T=0$.
In principle, for $d=2$ one should see a Kosterlitz-Thouless (KT) transition. However,
since our approximation does not treat the bosonic phase fluctuations in an adequate way,
we cannot expect to see the KT-scenario. 
On the other hand, $T_{C}$ may well be
finite for a  3-d system. Actually our procedure is similar in 
spirit to the two-particle self-consistent approach
to the repulsive Hubbard model by Vilk {\it et al.} \cite {vilk,vilk2}. \\
Now for a two-dimensional system, we can explicitly evaluate the number of pairs.
We obtain :
\begin{equation}\label{nb}
n_{d}={2R\over\beta WZ}\ln{\left( {e^{-\beta(W{Z\over 2}-\mu_{B})}-1\over
e^{\beta\mu_{B}}-1 }\right)}, \;\;\;\;\;\;\;\;\;\;\;\;\;\;\;\;\;\;\;\;\;\; 
\end{equation}
\begin{equation}\label{nf}
n_{F}={1\over\beta W\gamma}
\left(|E_{b}|+\Delta\over\Delta \right)\ln{\left( {e^{{\beta\over 2}\left( \mu_{B}-\Delta \right)}
+1\over e^{{\beta\over 2}\left( \mu_{B}-\Delta-\gamma W \right)}+1}\right)},
\end{equation}
where $\Delta$ and $\gamma$ have been given above. \\
We are now ready to (numerically) solve Eqns. (\ref{densto}), 
(\ref{nb}) and (\ref{nf}) for the chemical potential, 
as a function of temperature. Hereafter, the density $n$ is fixed to be $n=0.1$. The resulting
$\mu_{B}(T)$ vanishes exponentially at $T=0$. From Eqns. (\ref{g2}) 
and (\ref{t0}), 
we obtain that the range of the two-particle Green's function, $\xi$, 
is given by 
\begin{equation}
{\xi\over a}=\sqrt{{t\over 2|\mu_{B}|}}, 
\end{equation}
where $a$ is the lattice spacing. 
It is displayed on fig. 1, for three different values of the coupling 
strength $U$. 
We note that it shows a strong dependence on $U$. 
At very
low temperatures, it is given by :
\begin{eqnarray} 
{\xi\over a}=\sqrt{{\beta t\over2}}e^{{\beta WZn\over 8R}}.
\end{eqnarray} 
Oppositely, in the intermediate 
temperature range, $\xi$ becomes independent of $U$. 
We also find that $\xi$ decreases with increasing $|U|$.
In the strong coupling limit the ratio ${Z\over R}$ tends to zero and thus the exponential
divergence of $\xi$ in ${1 \over T}$ is suppressed (there remains only the power-law 
dependence $\xi\sim{1\over\sqrt{T}}$).\\
We may now define a coherence temperature
$T_{coh}$ as the temperature at which the range of the two-particle Green's function 
exceeds 10
lattice spacings. We obtain $T_{coh}\approx 0.16t$ for 
$U=-4t$ and $T_{coh}\approx 0.1t$ for 
$U=-6t$, which may be compared to $T_{C}$ as
obtained from the numerical simulations \cite{singer}. We see that they compare favorably and 
that, moreover, $T_{coh}$ decreases with increasing $U$.
Finally, we note that $\xi$ becomes of the order of the lattice spacing
at temperatures well below $|E_{b}|/2$. 
We now turn to the temperature dependence of the number of pairs. It is displayed on fig.
2, as a function of $T$, for three different values of $U$. At low temperatures, it is independent 
of $T$.
Since the binding energy of the pair (and the gap) 
decreases as  $U$ gets smaller, $n_{d}$ begins to decrease
at a lower $T$ for $U=-6t$ than for $U=-10t$, for example. In all cases $n_{d}$ decreases by
a factor 2 at $T\approx\Delta/2$.\\
In order to reach a better self-consistency, we now calculate how 
the quasi-bound states affect
the two-particle propagator. Introducing $G_{1}(\vec{q},\omega)$ as
the lower branch of 
$G(\vec{q},\omega)$ given 
by Eqn. (\ref{g2p}),
we can calculate the first correction to the two-particle propagator as $\chi_{1}$, with :
\begin{eqnarray}
\chi_{1}(\vec{q},i\nu_{n})=\sum_{i\omega_{n},\vec{p}}^{}{
G_{1}(\vec{q}-\vec{p},i\nu_{n}-i\omega_{n})
G_{0}(\vec{p},i\omega_{n})}\nonumber \\
\;\;\; +\sum_{i\omega_{n},\vec{p}}^{}{
G_{0}(\vec{q}-\vec{p},i\nu_{n}-i\omega_{n})G_{1}(\vec{p},i\omega_{n})}.
\end{eqnarray}
Carrying out the summation over Matsubara frequencies, we obtain that the 
statistical factors
are exponentially small and thus there is no correction to $\chi$ to that 
order. Thus, the T-matrix is still given
by Eqn. (\ref{t0}). We can now proceed to the calculation of the second-order correction to the
self-energy. It is given by :
\begin{equation}
\Sigma_{2}(\vec{q},i\omega_{n})=\sum_{i\nu_{n},\vec{p}}^{}{
T_{0}(\vec{p},i\nu_{n})G_{1}(\vec{p}+\vec{q},i\nu_{n}+i\omega_{n})},
\end{equation}
yielding, after performing analytical continuation,
\begin{eqnarray}
\Sigma_{2}(\vec{q},\omega)=\;\;\;\;\;\;\;\;\;\;\;\;\;\;\;\;\;\;\;\;\;\;\;\;
\;\;\;\;\;\;\;\;\;\;\;\;\;\;\;\;\;\;\;\;\;\;\;\;\;\;\;\;\;\;\;\;\;\;\;\;\;\;\;\;\;\;\;\;\;\;\;\nonumber
\end{eqnarray}
\begin{eqnarray}
\left( {|E_{b}|^{2}\Phi t\over W\gamma} \right)\left({x_{\vec{q}}-2\left( t_{\vec{q}}-
\mu \right)-\mu_{B}\over 2x_{\vec{q}}}\right)\nonumber \;\;\;\;\;\;\;\;\;\;
\;\;\;\;\;\;\;\;\;\;\;\;\;\;\;\;\;\;\;\;\;\;
\end{eqnarray}
\begin{eqnarray}
\times\left({1\over \omega-{1\over 2}\left( x_{\vec{q}}-\mu_{B} \right)+i0^{+}}\right)
\nonumber\;\;\;\;\;\;\;\;\;\;
\;\;\;\;\;\;\;\;\;\;\;\;\;\;\;\;\;\;\;\;\;\;\;\;\;\;\;\;\;\;\;\;\;\;\;
\end{eqnarray}
\begin{eqnarray}
+{U^{2}\omega_{c}\over W}
\left({x_{\vec{Q}-\vec{q}}-2\left(t_{\vec{Q}-\vec{q}}-\mu \right)-\mu_{B}\over 2x_{\vec{Q}-\vec{q}}}\right) \;\;\;\;\;\;\;\;\;\;\;\;\;\;\;\;\;\;\;\;\;\;\;\;\;\;\; \nonumber
\end{eqnarray}
\begin{eqnarray}\label{sig2}
\times\left({1\over \omega+2\mu+|U|-{1\over 2}\left(x_{\vec{Q}-\vec{q}}+\mu_{B}
\right)+i0^{+}}\right),\;\;\;\;\;\;\;\;\;\;\;\;
\end{eqnarray}
where $\omega_{c}$ is a frequency cut-off needed by the assumption that the
$\eta$-resonance is sharp for ${\left( \vec{Q}-\vec{q} \right)^{2}\over 4m^{*}}
\leq\omega_{c}
\approx {U^{2}\over W}$, with $m^{*}=m{|U|\over t}$. This corresponds to 
$|\vec{Q}-\vec{q}|{\ \lower-1.2pt\vbox{\hbox{\rlap{$<$}\lower5pt\vbox{\hbox{$\sim$}}}}\ }
{U \over W}$, as it is borne out by numerical calculations of the two-particle
Green's function $G^{(2)}$ covering the whole Brillouin zone \cite{morten,demler}. The precise
value of the cut-off has little influence in the following (numerical) results.
We note that the first contribution in Eqn. (\ref{sig2}) 
is following from the long-wavelength
behavior of the T-matrix, while the second is due to the $\eta$-resonance. Even though
both $\Sigma_{1}$ and the first contribution to $\Sigma_{2}$ 
originate from the pole of 
$T_{0}(\vec{q},i\nu_{n})$, given in Eqn. (\ref{t0}),
they are found to have opposite dispersions. The total self-energy results as :
\begin{eqnarray}\label{self1}
\Sigma(\vec{q},\omega)=\Sigma_{1}(\vec{q},\omega)+\Sigma_{2}(\vec{q},\omega).
\end{eqnarray}
At this point of the calculation, we can recalculate the particle density $n_{1}$
by evaluating explicitly Eqn. (\ref{densi}), using the Green's function resulting from
Eqn. (\ref{self1}). Since there is no a priori reason that this would
yield a result comparable to what is following from Eqn. (\ref{densto}), this is
a consistency check of the framework we are using. The result is displayed on 
fig. 3. It is obvious that this is in very good agreement with the expected value
$n=0.1$ for all values of $U$. This means that our calculation is consistent. It also implies
that, on one hand a small change in the chemical potential corresponds to
a small change in the particle density, but on
the other hand, it may induce a big change in the pair density.
\section*{(c) Excitation spectrum}
We can now summarize our findings by plotting the various
pieces of the spectral function. 
This is done on fig. 4a, for $U=-6t$ ($E_{b}=-t$) and on fig. 4b for $U=-10t$ ($E_{b}=-4.2t$) for momenta along the diagonal of the Brillouin zone.
In both cases, the spectrum consists of four branches. 
The spectral weights of the various branches are displayed on
fig. 5a for $U=-6t$ and fig. 5b for $U=-10t$. \\
These results can be commented as follows :\\
(i) There is a branch at
negative energies following from the quasi-bound states. It is centered around
$-{W\over 2}-{\Delta+|E_{b}|\over 4}$ with a width $W-{\Delta-|E_{b}|+2\mu_{B}\over 2}<W$.
The dependence in $U$ is weak, and the dispersions for $|U|=6t$ and
$|U|=10t$ are essentially the same, up to a small shift;
moreover, we note that they are opposite to the free fermion dispersion. 
The weight decreases with increasing momentum and gets mostly
negligible for $q_{x}\sim 2$. It decreases with increasing $|U|$, at small
$\vec{q}$ 's, but the total weight remains constant.\\ 
(ii) The fermionic band that was coming out of 
the first approximation (Eqn. (\ref{g2p})) is now
resulting as a superposition of two branches which would merge
into a single one, were damping 
taken into account. 
This superposition is done out of the two branches which are most free-electron-like.
This is somewhat arbitrary, especially in the domain where the hybridization with
the other branch, which lies at positive energies too (see below), is strong. 
Consequently, we do not show the spectral weights in this range.  
The width of the fermionic band is slightly larger than the original bandwidth, and shows
little dependence on $U$. It lies at slightly larger
energies for $|U|=10t$ than for $|U|=6t$, but again the dependence in $U$ is weak. 
Its dispersion is parallel to the free fermion one.\\
(iii) There is a 
third branch resulting from the $\eta$-
resonance; its dependence in $U$ is weak and both curves 
(for $|U|=6t$ and
$|U|=10t$) are parallel. It comes down in energy with increasing $|U|$.
The contribution of 
the $\eta$-resonance band is very small, barely
accounting for 5 percent of the spectral weight at its maximum, which is located 
at $\vec{q}=\vec{Q}$.\\
It is interesting to compare our results with Quantum Monte Carlo (QMC) simulations
\cite{singer,singer2}. The calculations of Ref.\cite{singer} have been performed for
$n=0.4$ and for $|U|=-4t, -8t, -12t$. This is likely to be outside the realm of densities 
for which the T-matrix approximation is really valid. Nevertheless, the following features coincide
with our findings : there are two distinct excitation branches which get more and more
clearly separated by a gap when $|U|$ increases. The width of the two bands is smaller
than in our calculation, in particular the lower (bosonic) band has very 
little dispersion. However, the weights (the fermionic band has large
weight for large wave vectors, whereas the weight of the negative energy 
bosonic band is concentrated near the zone center) is similar to our figures 4 and 5.\\
Ref.\cite{singer2} presents new results for the one-electron spectral function for
$n=0.1$, the same value that we have used, and $|U|=8t$. Here again, there is
a fermionic branch at positive energies $E$ (positive 
with respect to the chemical potential), separated
by a "pseudo-gap" from the excitations at $E<0$. Weight and width of the
fermionic band is similar to our result. The excitations at $E<0$ have most
weight near the zone center, as we find it. However, their structure seems to be more complex :
with increasing wave number they split into two "sub-branches". The lower part
has downward dispersion and seems to correspond to our bosonic branch. 
The other sub-branch produces weight near the chemical ($E\simeq 0$) potential for
large wave numbers. This might correspond to our $\eta$-resonance, provided that its energy
near the zone corner, lying at a positive energy in our calculation, would in reality be lower. In this respect, the spectral functions for
$n=0.4$ and $|U|=-6t$, also shown in Ref.\cite{singer2}, are particularly interesting :
for $E>0$, besides the strong fermionic branch, there is a second branch of excitations the
dispersions of which is very similar to the $\eta$-peak in our figures 4 and 5. \\
Thus, the main features of our spectral functions seem to be present in the Monte Carlo
results, although the latter show a more complex structure. A detailed and more
quantitative comparison with QMC should also take into account the fact the "Maximum
Entropy method" used there in order to extract spectral functions from data obtained as
functions of the (imaginary) Matsubara frequencies does not easily allow for an
unambiguous identification of excitations with small weight.\\
Finally, we note that we do not obtain
any spectral weight at zero frequency, signaling the presence of a correlation induced gap. 
We also checked that
including particle-hole excitations in the calculation does not affect this conclusion.
The appearance of a true gap in our calculation is, at least partly, due to the fact
that our spectral lines have no width (except that the "doubling" of the fermionic
branch gives a hint to a broadening of the latter). Spectral functions with finite
line width would be obtained either by doing the $\vec{q}$-sums in the
expression (\ref{s1tot}) 
for the self-energy more precisely, and/or by evaluating $G$ and $\Sigma$
by solving Eqns. (\ref{tfrai}) to (\ref{g1tot}) self-consistently.\\
\indent
In summary, we have determined the excitation spectrum of the attractive Hubbard
model at intermediate coupling out of a simple analytical calculation. We first pointed
out an intrinsic problem of perturbation theory relative
to the implementation of the Mermin-Wagner theorem. 
We made use of an alternative expression
for the density to obtain a qualitatively correct theory
which does not break down at low $T$ in two dimensions, in contrast to
previous self-consistent calculations. We obtain an analytical expression 
for the
Green's function which reproduces 
the qualitative features of the QMC simulations in the low 
density regime.\\
\indent
We acknowledge valuable discussions with
T. Schneider, J.M. Singer, M.H. Pedersen and J.J. 
Rodr\'{\i}guez-N\'{u}\~{n}ez. 
We thank J.M. Singer for providing us with his numerical data prior to 
publication. 
This work was 
financially supported by the Swiss National Science 
Foundation, under grants 20-43111.95 and 20-47149.96.  
M. Yu. K. acknowledges the University of Neuch\^{a}tel, where part of this 
work has
been performed, for hospitality and the Swiss National Science Foundation for 
partial funding.

\begin{figure}
\caption{Range of the two-particle Green's function, as a function of temperature $T$, 
at density $n=0.1$, for three different interaction strengths, $U=-4t$ (short-dashed line),
$U=-6t$ (full line), $U=-8t$ (dotted line) and 
$U=-10t$ (long-dashed line).}
\label{fig1}
\end{figure}
\begin{figure}
\caption{Density of quasi-bound pairs as a function of temperature $T$, 
at density $n=0.1$, for $U=-6t$ (full line), $U=-8t$ (dotted line) and 
$U=-10t$ (dashed line).}
\label{fig2}
\end{figure}
\begin{figure}
\caption{Particle density $n_{1}$ which is obtained by using Eqn. (25); the
chemical potential is calculated numerically out of Eqn. (32).}
\label{fig3}
\end{figure}
\begin{figure}
\caption{a) Spectrum of the Green's function at $n=0.1$, $T/t=0.1$ and
$U=-6t$ It consists of the bosonic band (dashed line), the
fermionic bands (dotted and dashed-dotted lines) and the 
$\eta$-resonance band (full line).
b) The same for $U=-10t$.}
\label{fig4}
\end{figure}
\begin{figure}
\caption{a) Spectral weight of the bosonic band (dashed line), the 
fermionic bands
(dotted line) and the $\eta$-resonance band (full line), for $n=0.1$, $T/t=0.1$ 
and
$U=-6t$. b) The same for $U=-10t$.}
\label{fig5}
\end{figure}

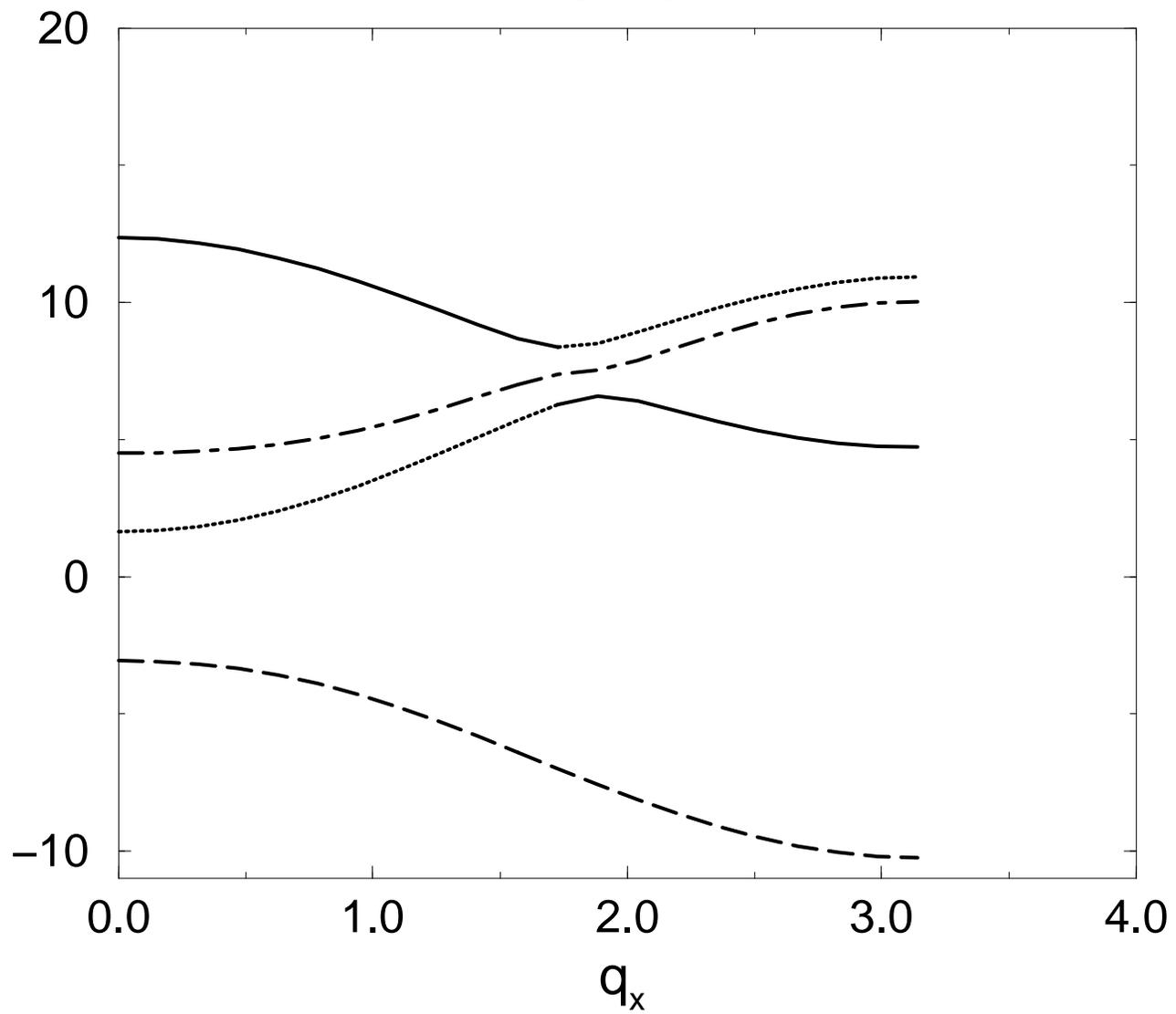
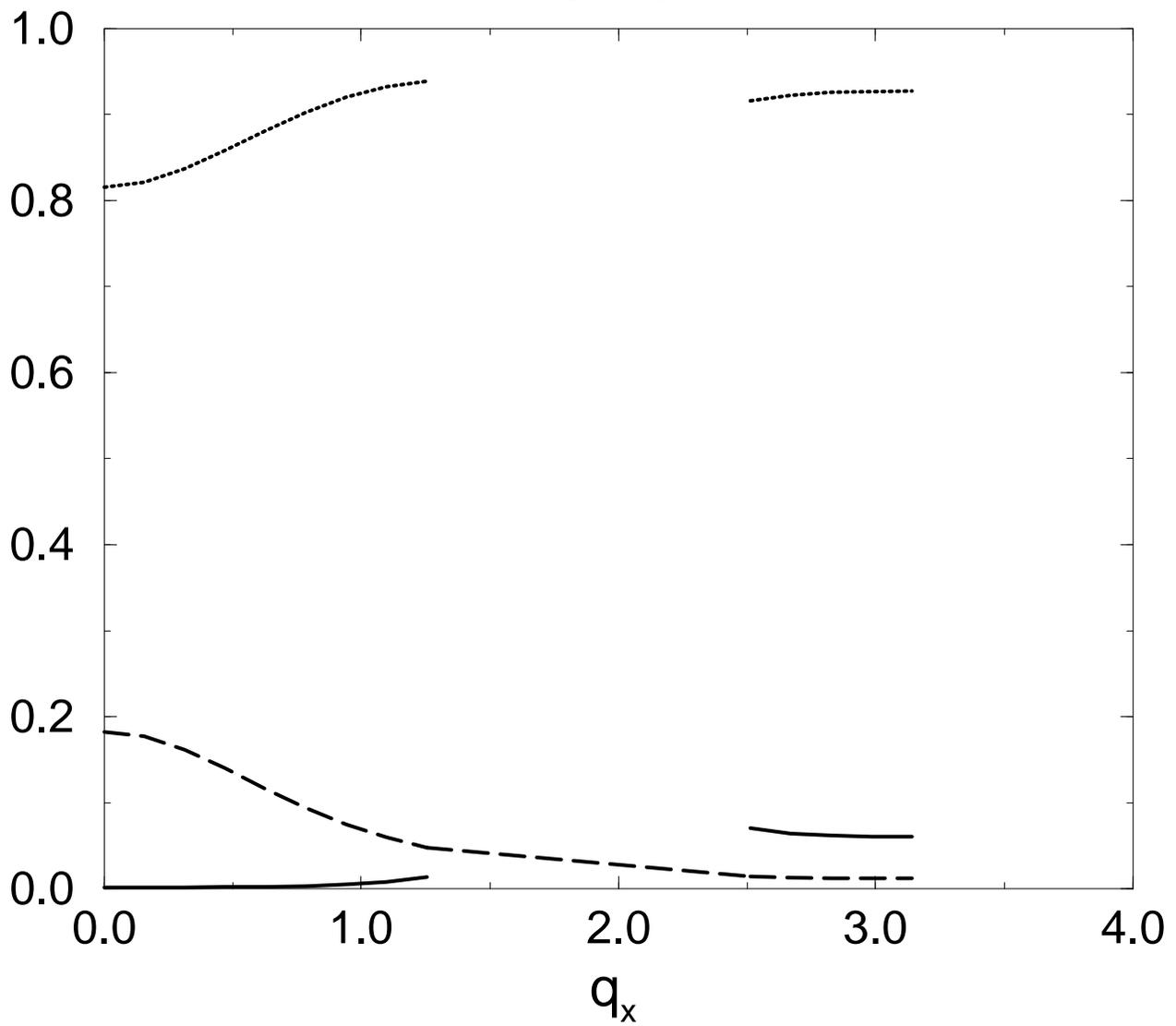
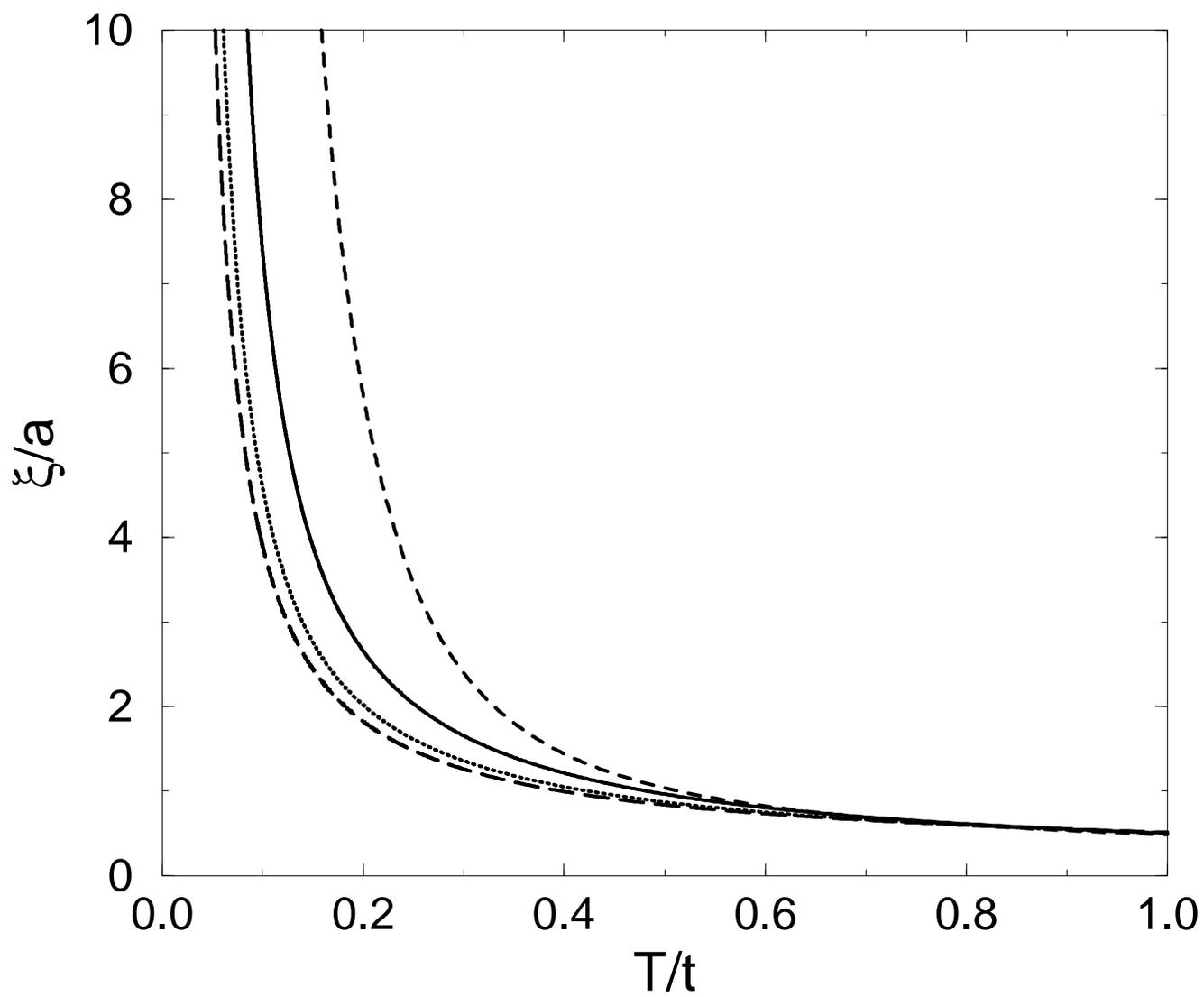
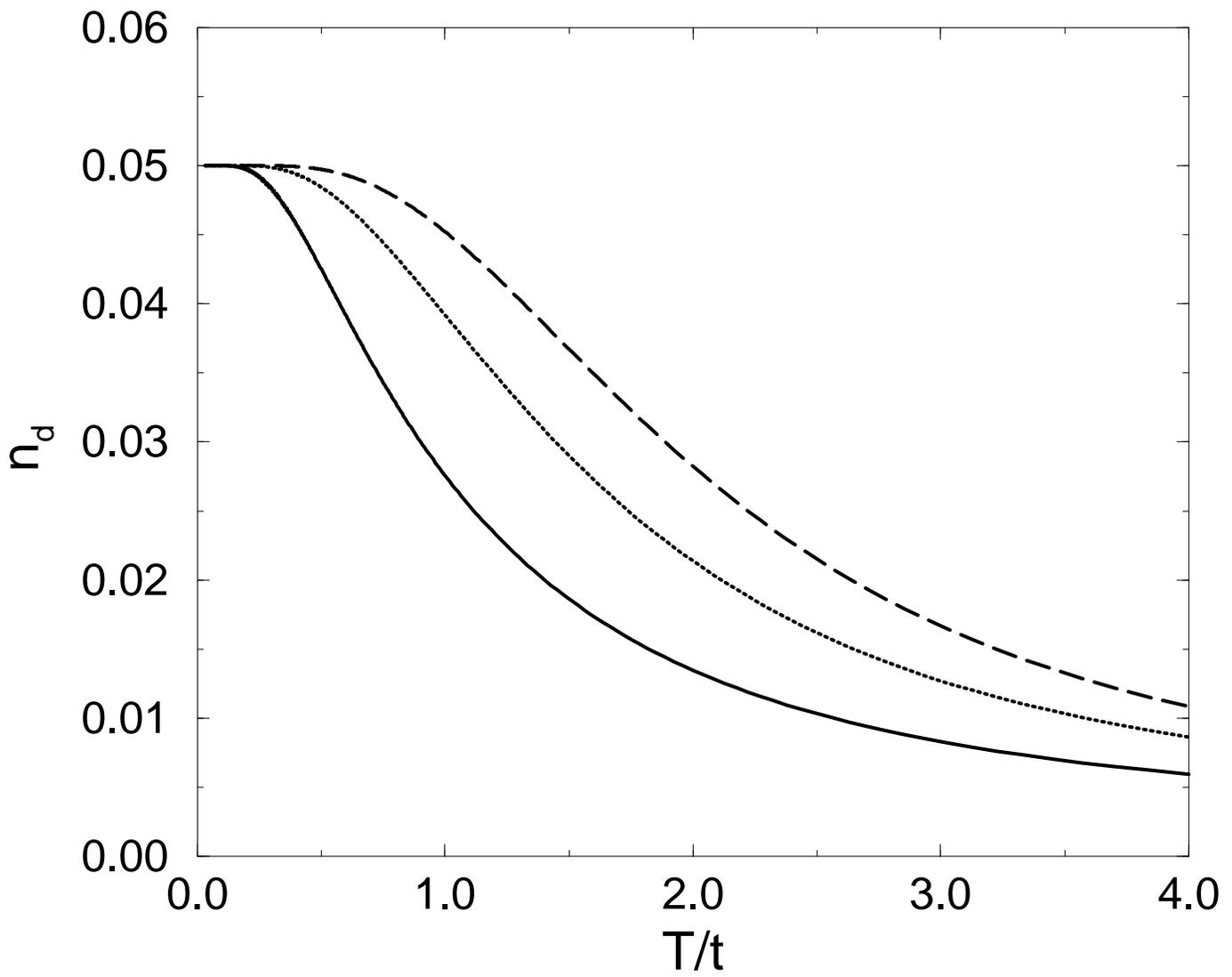
\end{document}